\documentclass[prl,aps,twocolumn,showpacs,floatfix, superscriptaddress]{revtex4}%
\usepackage{amsmath}
\usepackage{amsfonts}
\usepackage{amssymb}
\usepackage{graphicx}
\usepackage{bm}

\begin{document}

\title{Doping evolution of the absolute value of the London penetration depth and superfluid density in single crystals of Ba(Fe$_{1-x}$Co$_x$)$_2$As$_2$}
\author{R.~T.~Gordon}
\affiliation{Ames Laboratory, U.S. DOE, Ames, IA 50011}
\affiliation{Department of Physics \& Astronomy, Iowa State University, Ames, Iowa 50011}
\author{H.~Kim}
\affiliation{Ames Laboratory, U.S. DOE, Ames, IA 50011}
\affiliation{Department of Physics \& Astronomy, Iowa State University, Ames, Iowa 50011}
\author{N. Salovich}
\affiliation{Loomis Laboratory of Physics, University of Illinois at Urbana-Champaign, 1110 West Green St., Urbana, IL 61801}
\author{R.~W.~Giannetta}
\affiliation{Loomis Laboratory of Physics, University of Illinois at Urbana-Champaign, 1110 West Green St., Urbana, IL 61801}
\author{R.~M.~Fernandes}
\affiliation{Ames Laboratory, U.S. DOE, Ames, IA 50011}
\affiliation{Department of Physics \& Astronomy, Iowa State University, Ames, Iowa 50011}
\author{V.~G.~Kogan}
\affiliation{Ames Laboratory, U.S. DOE, Ames, IA 50011}
\author{T.~Prozorov}
\affiliation{Ames Laboratory, U.S. DOE, Ames, IA 50011}
\author{S.~L.~Bud'ko}
\affiliation{Ames Laboratory, U.S. DOE, Ames, IA 50011}
\affiliation{Department of Physics \& Astronomy, Iowa State University, Ames, Iowa 50011}
\author{P.~C.~Canfield}
\affiliation{Ames Laboratory, U.S. DOE, Ames, IA 50011}
\affiliation{Department of Physics \& Astronomy, Iowa State University, Ames, Iowa 50011}
\author{M.~A.~Tanatar}
\affiliation{Ames Laboratory, U.S. DOE, Ames, IA 50011}
\author{R.~Prozorov}
\email[corresponding author: ]{prozorov@ameslab.gov}
\affiliation{Ames Laboratory, U.S. DOE, Ames, IA 50011}
\affiliation{Department of Physics \& Astronomy, Iowa State University, Ames, Iowa 50011}

\date{10 June 2010}

\begin{abstract}
The zero temperature value of the in-plane London penetration depth, $\lambda_{ab}(0)$, has been measured in single crystals of Ba(Fe$_{1-x}$Co$_x$)$_2$As$_2$ as a function of the Co concentration, $x$, across both the underdoped and overdoped superconducting regions of the phase diagram.  For $x\gtrsim0.047$, $\lambda_{ab}(0)$ has been found to have values between 120 $\pm$ 50~nm and 300 $\pm$ 50~nm.  A pronounced increase in $\lambda_{ab}(0)$, to a value as high as 950 $\pm$ 50~nm, has been observed for $x\lesssim0.047$, corresponding to the region of the phase diagram where the itinerant antiferromagnetic and superconducting phases coexist and compete. Direct determination of the doping-dependent $\lambda_{ab}(0)$ has allowed us to track the evolution of the temperature-dependent superfluid density, from which we infer the development of a pronounced superconducting gap anisotropy at the edges of the superconducting dome.
\end{abstract}

\pacs{74.25.Nf,74.20.Rp,74.20.Mn}

\maketitle

\section{Introduction}

The zero temperature value of the London penetration depth is directly related to the superfluid density in the ground state of a system through $\lambda(0)\propto 1/\sqrt{n_s(0)}$ \cite{Tinkham_SC}. In the clean, low scattering limit, $n_s(0)$ is equal to the total density of conduction electrons, $n_N$. There are cases in which other phases, for example itinerant magnetism, can compete with superconductivity for the same conduction electrons, thus reducing the overall number of carriers in the superconducting state at $T=0$. Given the rich doping phase diagram of the newly discovered iron-based superconductors in which a long range magnetically ordered state, with itinerant character, coexists with a superconducting state, questions are raised regarding the effects of the competition between these states for the same electrons \cite{Canfield_TM, Ni2008, Pratt09, Christianson09, Drew2009, Laplace09, Goko09, Fernandes2009}. One way to approach this matter is to study the doping evolution of $\lambda_{ab}(0)$ across the phase diagram of these materials and use it to infer the corresponding change in the superfluid density, especially in the regime of the phase diagram where these two phases coexist. Determination of the absolute value of the London penetration depth is also important for the correct evaluation of the normalized, temperature-dependent superfluid density, $\rho_s(T)=\left(\lambda(0)/\lambda(T)\right)^2$. This quantity can be calculated from various models of the superconducting gap and provides insight into the pairing mechanism.

In the present study we focus on $\lambda_{ab}(0)$, which is the ground state screening length associated with supercurrent flowing in the crystallographic ab-plane as a result of an external magnetic field applied along the $c$-axis.  For $x \gtrsim0.047$, the measured values of $\lambda_{ab}(0)$ have been found between 120 $\pm$ 50~nm and 300 $\pm$ 50~nm.  A pronounced increase in $\lambda_{ab}(0)$ to a value as high as 950 $\pm$ 50~nm for $x\lesssim0.047$ has been observed. We interpret the increase in $\lambda_{ab}(0)$ for samples with $x\lesssim0.047$ to be due to the competition between the superconducting and itinerant antiferromagnetic states for the same conduction electrons.

%For T=0, the value of this quantity for a free electron gas is $\lambda(0)=mc^2/4\pi ne^2$.

The experimental determination of $\lambda(0)$ is a rather challenging task since only finite temperatures can be reached. There are techniques that are capable of obtaining an estimate of its value by taking advantage of the small variation of $\lambda(T)$ at low temperatures, which can be on the order of 1~nm/K, along with precision measurements.  One such technique is muon spin rotation ($\mu$SR) \cite{Sonier2007}, which has produced estimates for $\lambda_{ab}(0)$ of 320~nm in (Ba$_{1-x}$K$_x$)Fe$_2$As$_2$ ($T_c\simeq 32 K)$ \cite{Khasanov_BaK122, Evtushinsky2009}, 470~nm in (Ba$_{0.55}$K$_{0.45}$)Fe$_2$As$_2$ ($T_c\simeq$ 30 K) \cite{Aczel2008}, 230~nm in Ba(Fe$_{0.6}$K$_{0.4}$)$_2$As$_2$ ($T_c\simeq$ 38 K) \cite{Hiraishi},  250~nm in La(O$_{1-x}$F$_x$)FeAs \cite{Luetkens2009} and values ranging from 189~nm to 438~nm in the Ba(Fe$_{1-x}$Co$_{x}$)$_2$As$_2$ series \cite{Williams2009, Williams2010}.

Another technique, magnetic force microscopy, has reported $\lambda_{ab}(0)=325 \pm 50$~nm in Ba(Fe$_{0.95}$Co$_{0.05}$)$_2$As$_2$ \cite{Luan2010}.  In addition, optical reflectivity measurements have been used to estimate $\lambda_{ab}(0)$ in Ba(Fe$_{1-x}$Co$_{x}$)$_2$As$_2$ and reported values of $277 \pm 25$~nm for $x=0.06$ and $315 \pm 30$~nm for for $x=0.08$ \cite{Nakajima2010}.  It is important to compare the values of $\lambda(0)$ obtained by as many different techniques as possible because each experiment requires its own set of assumptions and modeling procedures.

Given the overall disparity between the measured values of $\lambda(0)$ from these different experimental techniques, it is valuable to perform a systematic study of $\lambda(0)$ as a function of doping in the series of which large, high quality single crystals having homogeneous doping levels are available, namely the Ba(Fe$_{1-x}$Co$_x$)$_2$As$_2$ series.  The samples used for this study were members of the Ba(Fe$_{1-x}$Co$_x$)$_2$As$_2$ series and were obtained from the same source as in Ref.~\cite{Ni2008} using the same growth procedure.

\begin{figure}[tb]
\includegraphics[width=1.0\linewidth]{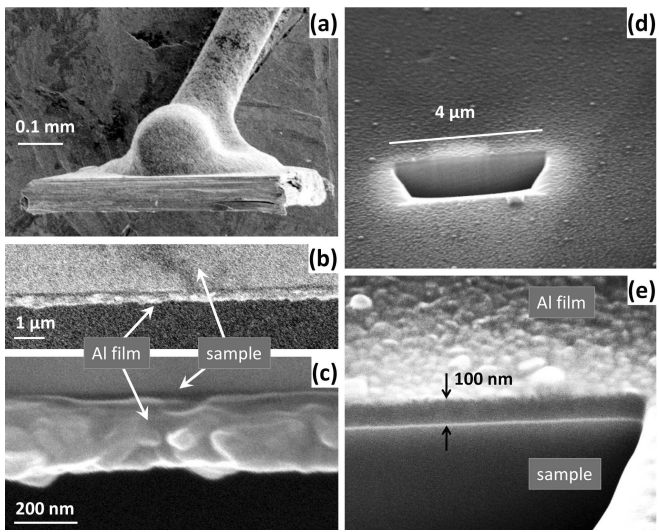}
\caption{Scanning electron microscope images of the Al coated samples. (a) Large scale view. The broken side is on top. (b) and (c) are zoomed in on the Al film on the edge of the broken side. (d) A trench produced by a focused-ion beam (FIB). (e) Close-up view of the FIB trench showing the Al film and its thickness.}
\label{fig1}
\end{figure}

\section{Experimental}

The experimental apparatus used for obtaining all of the penetration depth measurements in this work was a tunnel diode resonator (TDR) \cite{VanDegrift75}.  The essential components of the TDR are a tank circuit formed by an inductor and a capacitor, which has a resonance frequency $f_0=1/2\pi\sqrt{LC}\approx$ 14 MHz, and a tunnel diode.  While the diode is biased appropriately it serves as an $ac$ power source for the tank circuit.  To perform penetration depth measurements, the sample is mounted on a sapphire stage and inserted into the inductor coil.  The magnetic field of the coil, which is $\approx$ 10 mOe, is screened by the sample and thus  changes the inductance, L, and therefore also the resonance frequency by an amount $\Delta f$.  By utilizing $\Delta f(T)=-G4\pi\chi(T)=G[1-(\lambda(T)/R)\tanh(R/\lambda(T))]$, the TDR is capable of measuring the variation of the penetration depth in a superconductor, $\Delta\lambda(T)=\lambda(T)- \lambda(0)$, with a resolution of nearly 1 \AA, where $G$ is a geometry dependent calibration factor depending on the coil volume, sample volume, demagnetization and empty coil resonance frequency.  This calibration factor is measured directly by exctracting the sample from the inductor coil at its base temperature.

The TDR technique, as described above, provides very precise measurements of the variation of the penetration depth, $\Delta\lambda(T)$, but not the absolute value due to reasons described in detail in Ref.~\cite{Prozorov2000}. However, as proposed in the same reference, the TDR technique can be extended to obtain the absolute value of the penetration depth, $\lambda(T)$. The key to obtaining $\lambda(0)$ from TDR measurements is to coat the entire surface of the superconductor under study with a thin film of a conventional superconductor having a lower critical temperature and a known value of $\lambda(0)$. For this study, the aluminum films that were used to coat the Ba(Fe$_{1-x}$Co$_x$)$_2$As$_2$ samples had $T_c^{Al}\approx$ 1.2 K and thicknesses of 100~nm, as shown in Fig.~\ref{fig1}.

\begin{figure}[tb]
\includegraphics[width=1.0\linewidth]{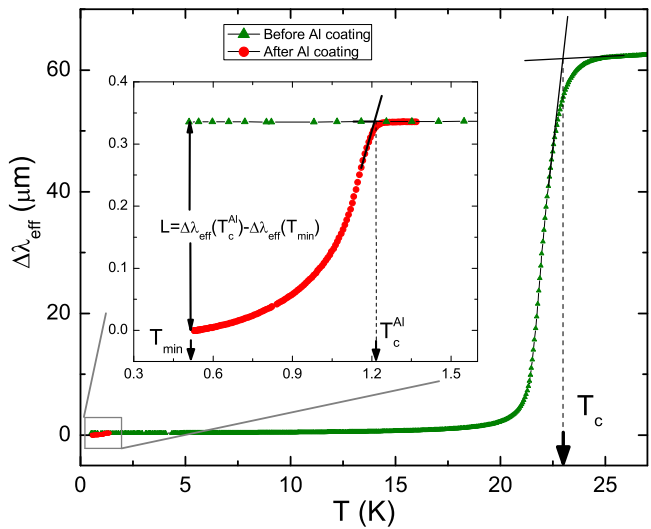}
\caption{(Color online) \underline{Main frame}: Full superconducting transition of an optimally doped Ba(Fe$_{0.93}$Co$_{0.07}$)$_2$As$_2$ crystal before and after coating. \underline{Inset}: Zoomed in low-temperature region, $T_{min} \lesssim T \lesssim T_c^{Al}$, before (green triangles) and after (red circles) the Al coating on the same sample. The overall frequency shift through the Al transition, denoted as $L$, is used for the calculation of $\lambda_{ab}(0)$.}
\label{fig2}
\end{figure}

 While the Al film is superconducting it participates with the coated superconductor to screen the magnetic field generated by the TDR coil. However, when it becomes normal it does effectively no screening because its thickness, $t$, is much less than the normal state skin depth at the TDR operating frequency of 14 MHz, where $\delta_{Al} \approx$ 75 $\mu$m for $\rho^{Al}_0$=10 $\mu \varOmega$-cm \cite{Hauser72}.  By measuring the frequency shift upon warming from $T_{min}$, which is the base temperature of the sample, to $T> T_c^{Al}$ we obtain the quantity $L\equiv\lambda_{eff}(T_c^{Al})-\lambda_{eff}(T_{min})$, shown in Fig.~\ref{fig2}. This quantity can be used to calculate $\lambda(0)$ along with the previously determined power-law relation for iron-based superconductors \cite{pair-break}, $\Delta\lambda(T) = \beta T^n$, and by using the formula for the effective magnetic penetration depth into both the Al film and the coated superconductor for $T< T_c^{Al}$, which is given by

\begin{equation}
\label{eqn1}
\lambda_{eff}(T)=\lambda_{Al}(T)\frac{\lambda(T)+\lambda_{Al}(T)\tanh{\frac{t}{\lambda_{Al}(T)}}}
{\lambda_{Al}(T)+\lambda(T)\tanh{\frac{t}{\lambda_{Al}(T)}}},
\end{equation}

\noindent where $\lambda(T)$ is the penetration depth of the coated superconductor and $\lambda_{Al}(T)$ is the penetration depth of the Al film.  As usual with the TDR technique, the variation of the penetration depth with temperature, $\Delta\lambda_{eff}(T)=\lambda_{eff}(T)-\lambda_{eff}(T_{min})$, is measured.  This method has been successfully demonstrated on several cuprate superconductors \cite{Prozorov2000} and has shown agreement with measurements of $\lambda(0)$ in Fe$_{1+y}$(Te$_{1-x}$Se$_x$) crystals obtained by different techniques \cite{Kim2010}. Here we use an extended analysis obtained by solving the appropriate boundary value problem.

The aluminum film was deposited onto each sample while it was suspended from a rotating stage by a fine wire in an argon atmosphere of a magnetron sputtering system. The formation of non-uniform regions in the film was avoided by bonding the wire to only a portion of the narrowest edge of each sample. Each film thickness was checked using a scanning electron microscope in two ways, both of which are shown in Fig.~\ref{fig1}.  The first method involved breaking a coated sample after all measurements had been performed to expose its cross section.  After this, it was mounted on an SEM sample holder using silver paste, shown in Fig.~\ref{fig1}(a). The images of the broken edge are shown for two different zoom levels in Fig.~\ref{fig1}(b) and (c). The second method used a focused ion beam (FIB) to make a trench on the surface of a coated sample, with the trench depth being much greater than the Al coating thickness, shown in Fig.~\ref{fig1}(d). The sample was then tilted and imaged by the SEM that is built into the FIB system, shown in Fig.~\ref{fig1}(e).

\section{Results and Discussion}

\begin{figure}[tb]
\includegraphics[width=1.0\linewidth]{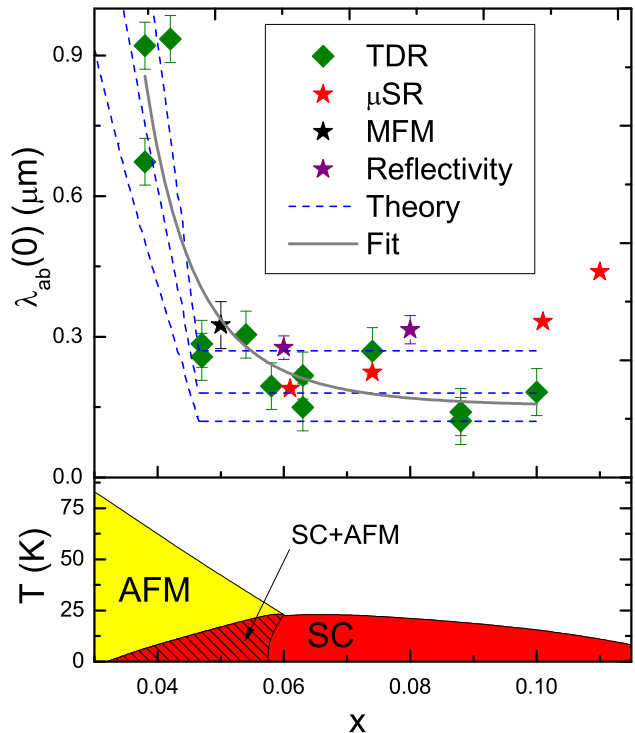}
\caption{(Color online) \underline{Top panel}: The zero temperature London penetration depth, $\lambda_{ab}(0)$, as a function of the Co concentration, $x$. The three dashed blue lines are theoretical curves obtained using Eq.~\ref{eqn2} for three different values of $\lambda_{ab}(0)$ in the pure superconducting state. The solid gray line is a guide to the eye (in the form of A+B/$x^n$).  Also shown are values of $\lambda_{ab}(0)$ obtained by other experiments for comparison explained in the text. \underline{Bottom panel}: Phase diagram for Ba(Fe$_{1-x}$Co$_x$)$_2$As$_2$ \cite{Canfield_TM, Ni2008, Fernandes2009, Nandi2010}. }
\label{fig3}
\end{figure}

The values of $\lambda_{ab}(0)$ that were obtained using the procedure described above for the Ba(Fe$_{1-x}$Co$_x$)$_2$As$_2$ system are shown in the top panel of Fig.~\ref{fig3} for doping levels, $x$, across the superconducting region of the phase diagram, shown schematically in the bottom panel of Fig.~\ref{fig3}. The size of the error bars for the $\lambda_{ab}(0)$ points was determined by considering the film thickness to be $t=100 \pm10$~nm and $\lambda_{Al}(0)=50\pm 10$~nm.  The discrepancy in $\lambda_{ab}(0)$ for the two samples having $x=0.038$ is clearly beyond these error bars and may possibly arise from cracks or inhomogeneities in the Al film, even though great care was taken to eliminate them during the coating process.  Thus, the error bars represent the uncertainty of the known parameters and the scatter in the data may arise from uncontrolled effects such as cracks or inhomogeneities in the aluminum films.  The scatter in the $\lambda_{ab}(0)$ values shown in the upper panel of Fig.~\ref{fig3} has an approximately constant value of 0.2 $\mu$m for all values of $x$, which probably indicates that the source of the scatter is the same for all samples.  For comparison, Fig.~\ref{fig3} also shows $\lambda_{ab}(0)$ obtained from $\mu$SR measurements (red stars) \cite{Williams2009, Williams2010}, the MFM technique (black star) \cite{Luan2010} and optical reflectivity (purple stars) \cite{Nakajima2010}, all in the Ba(Fe$_{1-x}$Co$_x$)$_2$As$_2$ system, most of which are consistent with our results within the scatter.  It may also be important to note that the $\lambda_{ab}(0)$ values from other experiments are all on the high side of the scatter that exists within the TDR $\lambda_{ab}(0)$ data set.  This is because any cracks or voids in the Al film will lead to underestimated values of $\lambda_{ab}(0)$. We note that we did not observe an increase in $\lambda_{ab}(0)$ towards the overdoped regime as reported from $\mu$SR measurements \cite{Williams2010}, although our values at the optimal doping do agree well.

Specifically, an increase in $\lambda_{ab}(0)$ on the underdoped side below $x \approx 0.047$ has been observed, which is in the region where the itinerant antiferromagnetic and superconducting phases coexist, as is shown in the bottom panel of Fig.~\ref{fig3}. The dependence of $\lambda_{ab}(0)$ on carrier concentration is $\lambda_{ab}(0)\propto 1/\sqrt{n_s(0)}$, where $n_s$ is the superfluid density, which is equal to the normal state carrier concentration in the clean case. The proportionality between $\lambda_{ab}(0)$ and $n_s(0)$ still holds if scattering is included, but $n_s$ is reduced due to a residual density of states within the gap. Overall, an increase in $\lambda_{ab}(0)$ is consistent with a decrease in the superfluid carrier concentration. There is compelling evidence suggesting that the itinerant antiferromagnetic spin density wave state in these materials acts to gap a portion of the Fermi surface \cite{Canfield_TM, Ni2008, Pratt09, Christianson09, Drew2009, Laplace09, Goko09, Fernandes2009}, which would remove mobile charge carriers and this qualitative idea is consistent with our experimental observations of the doping dependence of $\lambda_{ab}(0)$.  Changes in the Hall coefficient for these materials, moving from the pure superconducting region to the coexistence region, have also been interpreted as being due to the interaction between these phases \cite{Mun09, Fang09}.  It has been shown that the opening of a superconducting gap in the antiferromagnetic state transfers optical spectral weight from a mid-infrared peak to a Drude peak, even when the reconstructed Fermi surface would be fully gapped \cite{RMF_Drude}.  As a result, the coexistence state has a finite $n_s$, although smaller than in the pure superconducting state.

In order to provide a more quantitative explanation for the observed increase in $\lambda_{ab}(0)$ as $x$ decreases in the underdoped region, we have considered the case of $s^{\pm}$ superconductivity coexisting with itinerant antiferromagnetism \cite{Fernandes2009}.  For the case of particle hole symmetry (nested bands), the zero temperature value of the in-plane penetration depth in the region where the two phases coexist is

\begin{equation}
\label{eqn2}
\lambda^{SC+SDW}_{ab}(0)=\lambda^0_{ab}(0)\sqrt{1+\frac{\Delta^2_{AF}}{\Delta^2_0}}
\end{equation}

\noindent where $\lambda^0_{ab}(0)$ is the value for a pure superconducting system with no magnetism present, and $\Delta_{AF}$ and $\Delta_0$ are the zero temperature values of the antiferromagnetic and superconducting gaps, respectively.  Deviations from particle hole symmetry lead to a smaller increase in $\lambda^{SC+SDW}_{ab}(0)$, making the result in Eqn.~\ref{eqn2} an upper estimate.  For more information on the details of the calculation and the values of $\Delta_{AF}$ and $\Delta_0$ used, see Ref.~\cite{Fernandes2009, RMF_Drude}.

The three blue dashed lines shown in the top panel of Fig.~\ref{fig3}, which were produced using Eqn.~\ref{eqn2}, show the expected increase in $\lambda_{ab}(0)$ in the region of coexisting phases below $x\approx0.047$ by normalizing to three different values of $\lambda_{ab}(0)$ in the pure superconducting state, with those being 120~nm, 180~nm and 270~nm.  This theory does not take into account changes in the pure superconducting state, so for $x>0.047$ the dashed blue lines are horizontal.  These theoretical curves were produced using parameters that agree with the phase diagram in the bottom panel of Fig.~\ref{fig3} \cite{Fernandes2009, Nandi2010}, which includes a shift of the coexistence region to lower values of $x$ by an amount of 0.012, and given the simplifications of the model, the agreement with the experimental observations is quite reasonable.

While the exact functional form of $\lambda(x)$ is unknown, the solid gray line in Fig.~\ref{fig3} serves as a useful guide to the eye (of the form A+B/$x^n$), which does indeed illustrate a dramatic increase of $\lambda_{ab}(0)$ in the coexistence region and also a relatively slight change in the pure superconducting phase.
It should be noted that a dramatic increase in $\lambda_{ab}(0)$ below $x\approx0.047$ cannot be explained by impurity scattering, which would only lead to relatively small corrections in the magnitude of $\lambda_{ab}(0)$ (but, indeed, significantly affects the temperature dependence of $\lambda(T)$ \cite{Gordon2010}).

\begin{figure}[tb]
\includegraphics[width=1.0\linewidth]{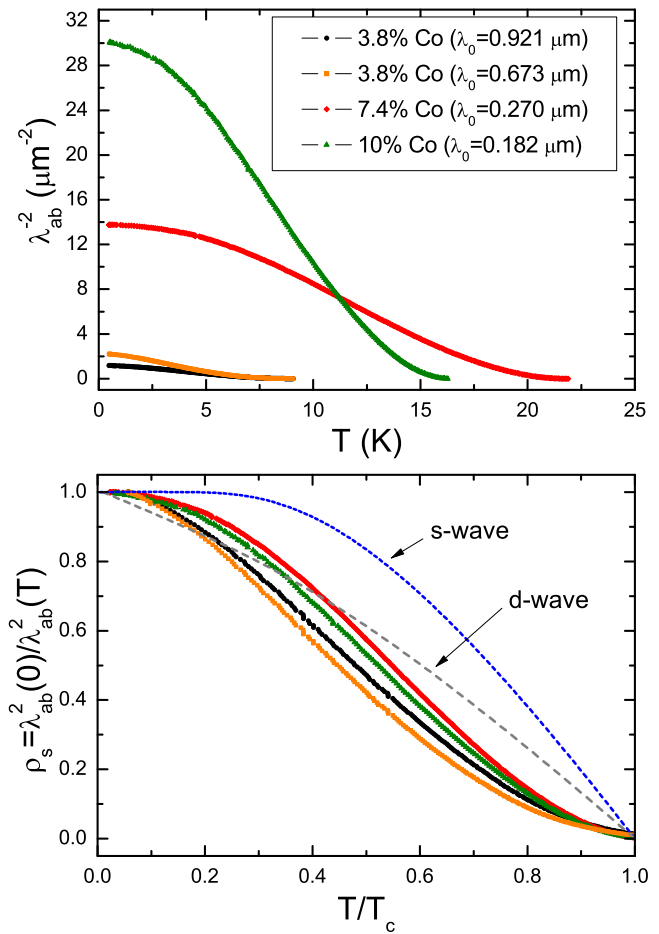}
\caption{(Color online) \underline{Top panel}: $\lambda_{ab}^{-2}(T)$ for samples of different Co concentrations, $x$, from the Ba(Fe$_{1-x}$Co$_x$)$_2$As$_2$ series constructed from previously measured $\Delta\lambda_{ab}(T)$ curves and $\lambda_{ab}(0)$ from this study. \underline{Bottom panel}: Normalized superfluid density, $\rho_s(T)$, for the same samples shown in the top panel along with the standard s-wave and d-wave cases for low impurity scattering.}
\label{fig4}
\end{figure}

Values of $\lambda_{ab}(0)$ obtained here can be used to calculate the actual penetration depth, $\lambda_{ab}(T)=\Delta\lambda_{ab}(T)+\lambda_{ab}(0)$, where $\Delta\lambda_{ab}(T)$ has been measured for each Ba(Fe$_{1-x}$Co$_x$)$_2$As$_2$ crystal used in this study before Al coating \cite{RTGPRL09, RTGPRBR09}. In the top panel of Fig.~\ref{fig4}, we examine $\lambda^{-2}_{ab}(T)\propto n_s(T)/m^*$ as a function of temperature in four different samples; with $x=0.038$ corresponding to the two far underdoped samples having different measured values of $\lambda_{ab}(0)$ in the region of coexisting phases, $x=0.074$ being close to optimal doping and $x=0.10$ being an overdoped concentration, all of which were used to determine the values of $\lambda_{ab}(0)$ shown in the top panel of Fig.~\ref{fig3}.  It should be noted that the orange and black curves in Fig.~\ref{fig4} for $x=0.038$ were made using the same $\Delta\lambda_{ab}(T)$ data, but different values of $\lambda_{ab}(0)$ because the temperature dependence of only one of the two samples shown in the top panel of Fig.~\ref{fig3}  was measured before aluminum coating.  As can be seen in the top panel of Fig.~\ref{fig4}, the values  of $\lambda^{-2}_{ab}(T\to0)$ are quite large for $x\gtrsim0.047$ (red diamonds and green triangles) relative to those with $x\lesssim0.047$ because the measured values of $\lambda_{ab}(0)$ are much smaller than 1 $\mu$m, i.e. 0.182~nm and 0.270~nm.  However, for $x\lesssim0.047$ (black circles and orange squares) the values of $\lambda^{-2}_{ab}(0)$ vary much less because $\lambda_{ab}(0)$ becomes closer to 1 $\mu$m, i.e. 0.673~nm and 0.921~nm.

Using the same penetration depth data that was used in the top panel of Fig.~\ref{fig4}, we construct the normalized superfluid density (phase stiffness), $\rho_s(T)=\left(\lambda(0)/\lambda(T)\right)^2$, which is commonly used to analyze penetration depth data and a quantity which is fairly easy to calculate for an arbitrary gap structure. The bottom panel in Fig.~\ref{fig4} shows $\rho_s(T)$ for the same samples shown in the top panel. The black and orange curves were constructed using the same $\Delta\lambda_{ab}(T)$, but different $\lambda_{ab}(0)$ values for the heavily underdoped sample, $x$=0.038, and the red and green curves are the data for optimally doped and overdoped compositions, respectively. Also shown for comparison are the $\rho_s(T)$ curves for a single band s-wave superconductor (dotted blue line) and a d-wave superconductor (dotted gray line), both in the clean limit. From Fig.~\ref{fig4}, $\rho_s(T\to0)$ and $\rho_s(T\to T_c)$ behave quite differently for the members of the Ba(Fe$_{1-x}$Co$_x$)$_2$As$_2$ series compared to the standard, single gap s-wave and d-wave clean limit cases. Impurity scattering would turn the d-wave curve quadratic at low temperatures, while leaving s-wave almost intact.

The data for all doping levels show an overall similar trend of the evolution of $\rho_s(T)$ across the phase diagram.
A special feature of this behavior is the negative curvature just below $T_c$. This behaivor suggests that below $T_c$ the superconducting gap develops slower than it does in the case of a single gap, which is a feature consistent with the behavior of $\rho_s(T)$ in a two-gap superconductor \cite{Kogan2gap}. Furthermore, the normalized $\rho_s(T)$ curve for the optimally doped sample over the entire temperature range stays above the curves for both heavily underdoped and overdoped samples. This distinction between the different Co-doping compositions suggests that the gap anisotropy, which is generally considered as being either the actual angular variation in $k-$space and/or the development of an imbalance between the gaps on different sheets of the Fermi surface, notably increases in the overdoped and underdoped compositions. This is consistent with the measurements of the specific heat jump \cite{Budkoheatcapacity} and the residual term \cite{Goffryk}, as well as with measurements of thermal conductivity \cite{Tanatar,Reid}. Thermal conductivity measurements with heat flow along the $c$-axis actually suggest that nodal regions develop in the superconducting gap in heavily under- and over-doped compositions. Indeed, measurements of $\lambda_{c}$ in a closely related Ba(Fe$_{1-x}$Ni$_x$)$_2$As$_2$, also suggest the existence of nodes in the superconducting gap \cite{Martin-Ni-doping}.

\section{Conclusion}

In conclusion, the zero temperature value of the in-plane London penetration depth, $\lambda_{ab}(0)$, has been measured for the Ba(Fe$_{1-x}$Co$_x$)$_2$As$_2$ series across the superconducting ``dome" of the phase diagram using an Al coating technique along with TDR measurements.  There is a clear increase in $\lambda_{ab}(0)$ below $x\approx 0.047$, which is consistent with a reduction in the superfluid density due to the competition between itinerant antiferromagnetism and superconductivity for the same electrons. The measured values of $\lambda_{ab}(0)$ were also used to construct the normalized superfluid density (phase stiffness), $\rho_s(T)$, and study its evolution with doping.  The negative curvature of $\rho_s(T)$ just below $T_c$ for samples across the superconducting dome of the phase diagram implies two-gap superconductivity. A notable suppression of $\rho_s$ for heavily underdoped and slightly overdoped samples with respect to samples with optimal doping suggests a developing anisotropy of the superconducting gap toward the edges of the superconducting dome, consistent with the behavior found in specific heat and thermal conductivity studies.

We thank J. Schmalian and A. Kreyssig for useful discussions. Work at the Ames Laboratory was supported by the division of Materials Science and Engineering, Basic Energy Sciences, Department of Energy (USDOE), under Contract No. DEAC02-07CH11358. Work at UIUC was supported by the Center for Emergent Superconductivity, an Energy Frontier Research Center funded by the USDOE Office of Science, Basic Energy Sciences under Award Number DE-AC0298CH1088. R.P. acknowledges support from the Alfred P. Sloan Foundation.

\end{document}